\documentclass[twocolumn]{emulateapj}
\usepackage{amsmath}
\usepackage{lipsum}
\usepackage{graphicx}
\shorttitle{Red Clump Stars in GALEX and Gaia}
\shortauthors{MOHAMMED ET AL.}
\usepackage{hyperref}

\def\deg{\ifmmode^\circ\else$^\circ$\fi} 

\begin{document}

\title{An Ultraviolet-Optical Color-Metallicity relation for Red Clump Stars using GALEX and Gaia}

\author{Steven Mohammed}
\affiliation{Department of Astronomy, Columbia University, New York, NY 10027}

\author{David Schiminovich}
\affiliation{Department of Astronomy, Columbia University, New York, NY 10027} 

\author{Keith Hawkins}
\affiliation{Department of Astronomy, Columbia University, New York, NY 10027}
\affiliation{Department of Astronomy, The University of Texas at Austin, 2515 Speedway Boulevard Stop C1400, Austin, TX 78712}

\author{Benjamin Johnson}
\affiliation{Harvard-Smithsonian Center for Astrophysics, Cambridge, MA 02138}

\author{Dun Wang}
\affiliation{Center for Cosmology and Particle Physics, Department of Physics, New York University, New York, NY 10003}, 

\author{David W. Hogg}
\affiliation{Center for Cosmology and Particle Physics, Department of Physics, New York University, New York, NY 10003}

\email{smohammed@astro.columbia.edu}

\begin{abstract}
Although core helium-burning red clump (RC) stars are faint at ultraviolet wavelengths, their ultraviolet-optical color is a unique and accessible probe of their physical properties. Using data from the GALEX All Sky Imaging Survey, Gaia Data Release 2 and the SDSS APOGEE DR14 survey, we find that spectroscopic metallicity is strongly correlated with the location of an RC star in the UV-optical color magnitude diagram.  The RC has a wide spread in (NUV - G)$_0$ color, over 4 magnitudes, compared to a 0.7-magnitude range in (G$_{BP}$ - G$_{RP}$)$_0$. We propose a photometric, dust-corrected, ultraviolet-optical (NUV - G)$_0$ color-metallicity [Fe/H] relation using a sample of 5,175 RC stars from APOGEE. We show that this relation has a scatter of 0.28 dex and is easier to obtain for large, wide-field samples than spectroscopic metallicities. Importantly, the effect may be comparable to the spread in RC color attributed to extinction in other studies.
\end{abstract}
 

\section{Introduction}
The Milky Way is host to a variety of stars spanning the entire stellar lifetime range. Average stars like our Sun eventually become red giants, some of which populate a prominent feature in color-magnitude diagrams (CMDs) called the red clump (RC) consisting of low-mass, metal-rich stars in the core helium-burning stage of stellar evolution. A sizable fraction of Solar neighborhood giants observed with Hipparcos are RC candidates (60$\%$, \citealt{girardi16}). Metallicities for these stars are readily available from surveys such as APOGEE. The metallicity of RC stars can be used to understand the star formation history and ages of stars in the Milky Way and inform stellar evolutionary models of RC, red giant and horizontal branch stars (\citealt{girardi16}).

With the release of Gaia DR2 (\citealt{gaia}), the Milky Way can now be probed to greater depths than ever before. The Gaia DR2 release presents parallax measurements for over 1 billion stars, which provide crucial distance information and G-band measurements to allow construction of the CMD of the Milky Way field population. We combine NUV-band data from the GALEX All Sky Imaging Survey (GAIS, \citealt{galex}) with Gaia and a catalog of RC stars from APOGEE DR14 (\citealt{ting18}). RC stars are very faint in NUV and should separate clearly from the main sequence. As RC stars have not been extensively probed in UV, this study has the potential to strengthen our understanding the relation between UV-optical colors and the physical properties of stars in this core helium-burning stage. In this paper, we focus on a (NUV - G)$_0$ color-metallicity relation and show how it compares to a similar color-metallicity relation derived using optical colors. Finally, we also discuss how this relation compares to predictions from stellar evolutionary model tracks, using the MIST code (\citealt{mist}, \citealt{choi16}, \citealt{paxton11}, \citealt{paxton13}, \citealt{paxton15}).

\begin{figure*}[]
\centering
\includegraphics[width=1\textwidth]{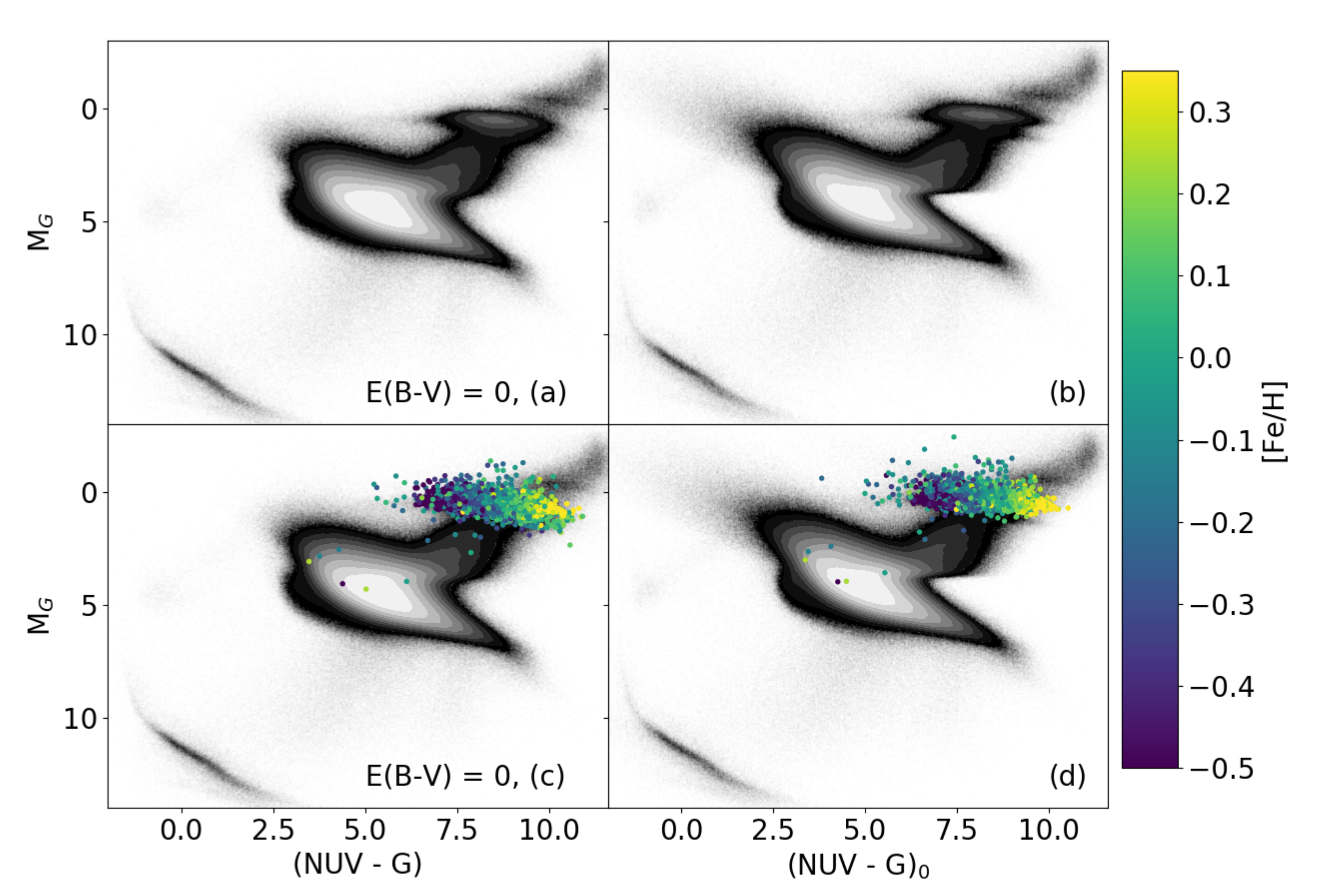}
\caption{M$_G$ vs (NUV - G)$_0$ distribution for the full parallax-selected ($<3.5$ kpc) GAIS-Gaia catalog  (top panels) with matches with RC stars shown (bottom panels). The left panels show the CMD without a Galactic extinction correction and the right panels apply a correction as described in the text. The main locus at the center of each panel shows the main sequence.  The RC stars are overlaid and colored by [Fe/H]. There is a clear trend of metallicity with NUV - G color in both cases. [Fe/H] from APOGEE DR14.}
\end{figure*}

\begin{figure*}[]
\centering
\includegraphics[width=1\textwidth]{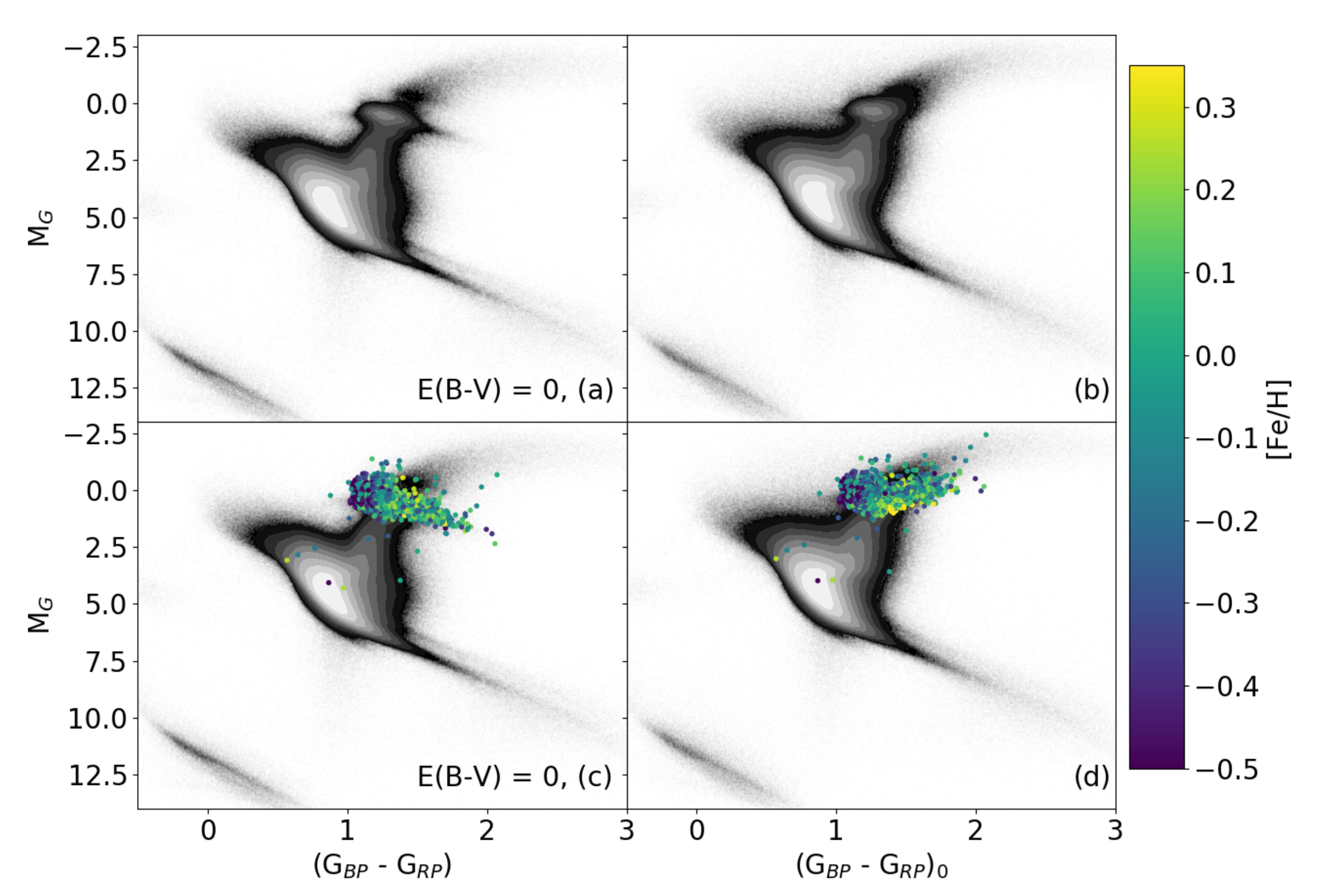}
\caption{M$_G$ vs G$_{BP}$ - G$_{RP}$ distribution for the full GAIS-Gaia catalog (top panels) with matches with RC stars shown (bottom panels). The left panels show the CMD without a Galactic extinction correction and the right panels apply a correction as described in the text. }
\end{figure*}

\section{Observations}
\subsection{Red Clump data}\affiliation{Department of Astronomy, The University of Texas at Austin, 2515 Speedway Boulevard Stop C1400, Austin, TX 78712, USA}

We build a sample of 5,175 RC stars from \citealt{ting18} which is constructed using data from the APOGEE (Apache Point Observatory Galactic Evolution Experiment, \citealt{apogee2017}) and LAMOST (Large Sky Area Multi-Object Fibre Spectroscopic Telescope, \citealt{lamost}) surveys. \citealt{ting18} build a RC sample of 92,249 Milky Way stars from APOGEE DR14 and LAMOST DR3 data with $\sim$ 3$\%$ contamination from red giant stars. For our analysis we only used the pristine RC sample obtained from APOGEE spectra. From its high signal-to-noise, near-infrared (1.51-1.70 $\mu$m) spectra, derived parameters such as metallicities, T$_{\rm{eff}}$, log g parameters are available, as well as abundances for many elements. APOGEE elemental abundances are typically accurate to 0.2 dex over the metallicity range considered here (\citealt{ASPCAP}).

\subsection{GAIS and Gaia DR2}
The GALEX All Sky Imaging Survey (GAIS) contains NUV data for millions ofx objects across the entire sky. Gaia DR2 provides Gaia G, G$_{BP}$, and G$_{RP}$ magnitudes and parallaxes that can be used to obtain distance information. The errors in G, G$_{BP}$, and G$_{RP}$ are of the order of millimagnitudes. We apply a small error-dependent correction to the Gaia parallaxes (\citealt{lutzkelker73}, \citealt{Oudmaijer98}). We then invert the parallax to get a distance. To cross-match these data we use the astropy (\citealt{astropy}) function \textit{search around sky} with a search radius of 3 arcseconds. In total we utilize coverage in GALEX NUV, Gaia G, G$_{BP}$, G$_{RP}$ and the relevant APOGEE footprint.

Galactic extinction plays a much larger role for the NUV than the other bands (\citealt{ccm89}) and will have a nontrivial effect on the location of objects in a CMD. To account for this reddening we use the 3D dust map from \citealt{GSF15}, which gives E$_{B - V}$ as a function of distance, in conjunction with Gaia parallax-derived distances to estimate the reddening in the line of sight of each object in this catalog. The NUV - G color is dust-corrected (indicated by a 0 subscript) using these E$_{B - V}$ values, adopting R$_{NUV}$ from \citealt{yuan13}, and R$_G$ from \citealt{jordi2010} who obtain R$_G$ values between 2.4 and 3.6. For our analysis, the extinction corrections for NUV and G are NUV$_0$ = NUV - E$_{B - V}$ $\times$ 7.24 and G$_0$ = G - E$_{B - V}$ $\times$ 2.85. We are restricted to the sky coverage of the \citealt{GSF15} map and remove any objects in the GAIS-Gaia catalog that do not overlap with the map. 

For the final catalog we make several additional cuts to the data. The final catalog contains objects that have detections in NUV, G, G$_{BP}$, and G$_{RP}$, [Fe/H] and T$_{\rm{eff}}$ measurements, parallax errors less than 10$\%$, $visibility$\textunderscore$periods$\textunderscore$used$ $>$ 8, and distances less than 3500 pc. Additionally, we use the $RC$\textunderscore$Pristine$ Classification from \citealt{ting18}. We do not require the APOGEE flags to be set to 0 for these objects (65 in total) however the removal of these objects do not impact our results. Our final catalog of GAIS and Gaia objects contains 10,357,542 objects. We cross-match this catalog with the RC catalog and obtain 5,175 matches. 

GAIS-Gaia does not appear to be limited to only very blue objects despite the expectation that GALEX would not observe many red stars. There is a large population at the expected position of the RC in the CMD. There are about 91$\%$ of objects in the main catalog along the Main Sequence versus 4$\%$ of objects in the RC. 

\begin{figure}[]
\centering
\includegraphics[width=0.5\textwidth]{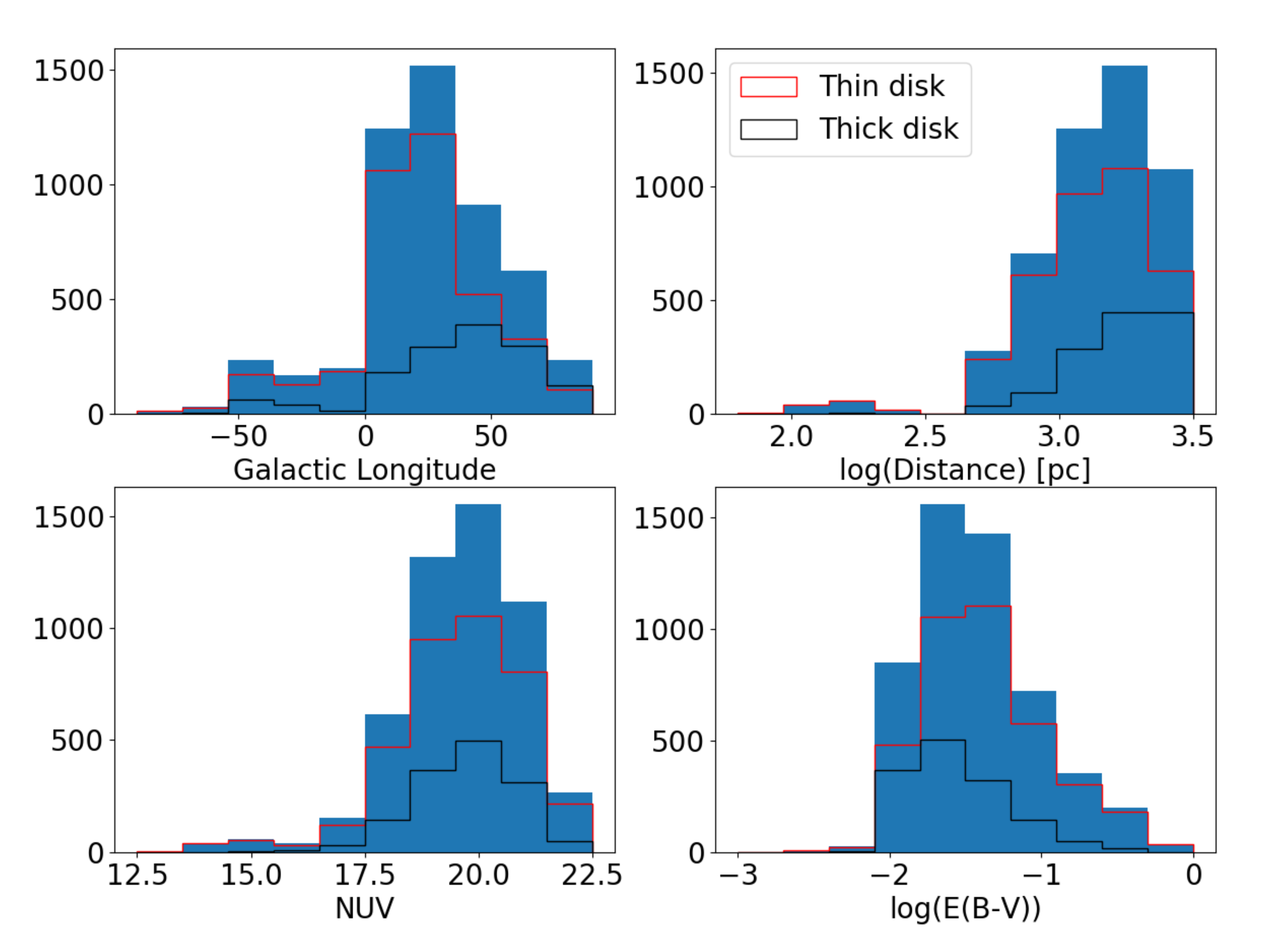}
\caption{Various statistics for our GAIS-Gaia RC sample. The sample is mostly restricted to the upper Galactic plane and at distances greater than 500 pc. We also overplot the thin and thick disk populations discussed further below.}
\end{figure}

\begin{figure}[]
\centering
\includegraphics[width=0.5\textwidth]{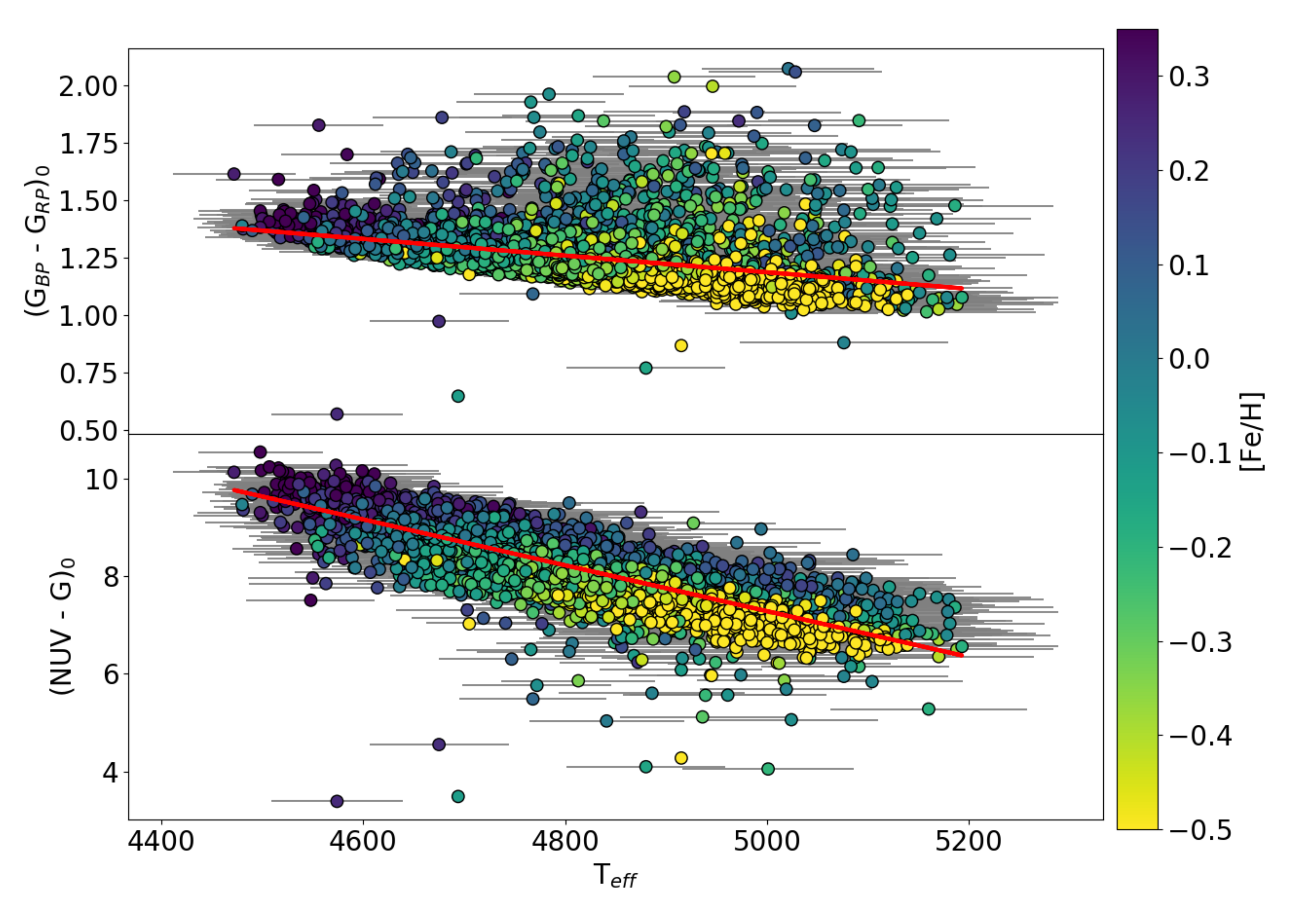}
\caption{T$_{\rm{eff}}$ from APOGEE vs extinction corrected (G$_{BP}$ - G$_{RP}$)$_0$ and (NUV - G)$_0$ colored by [Fe/H] for our RC sample. (G$_{BP}$ - G$_{RP}$)$_0$ shows significant scatter.  (NUV - G)$_0$ is more tightly correlated with T$_{\rm{eff}}$, although we also identify a subpopulation of very blue outliers.}
\end{figure}

\begin{figure}[]
\centering
\includegraphics[width=0.5\textwidth]{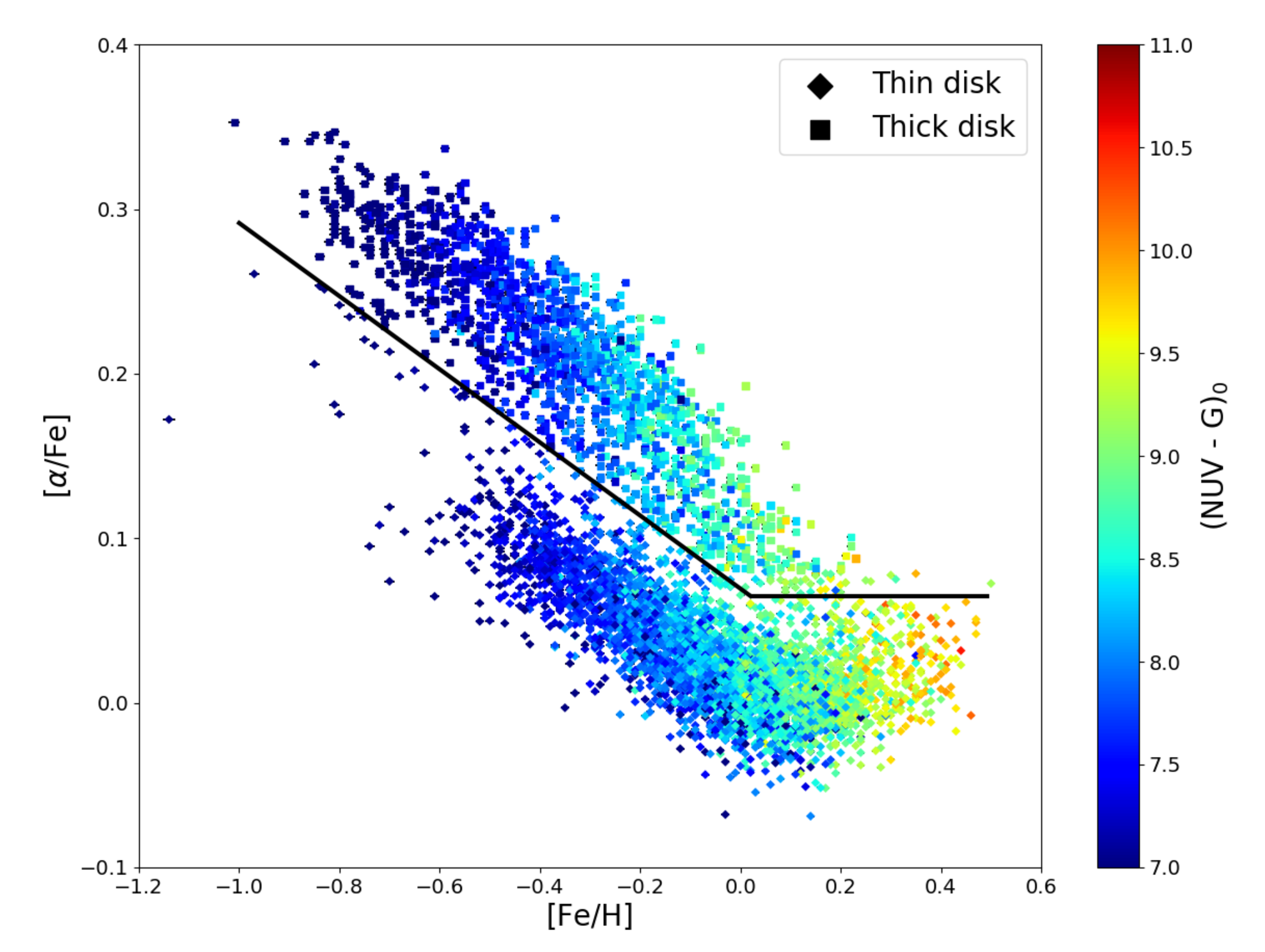}
\caption{[$\alpha$/Fe] vs [Fe/H] colored by the dust corrected (NUV - G)$_0$ (described in Figure 1). \citealt{hawkins15} discuss a cartoon depiction of this trend separating the lower branch (thin disk stars) from the upper branch (thick disk stars).}
\end{figure}

\section{Results}
Figures 1 and 2 show optical and UV-optical CMD histograms for the GAIS-Gaia catalog, using both uncorrected and Galactic extinction-corrected magnitudes. The general shape of the UV-optical CMD is very similar to that of optical: a large main sequence with the red giant branch and RC prominently displayed. The main sequence stretches from (NUV - G)$_0$ = 8 and M$_G$ = 6 to (NUV - G)$_0$ = 2.5 and M$_G$ = 1, and is where the bulk of the survey matches appear. The secondary locus around (NUV - G)$_0$ = 8 and M$_G$ = 0 (\citealt{hawkins17}) is populated by red giants, notably RC stars. The spread of the entire RC in (NUV - G)$_0$ is unlike that seen in (G$_{BP}$ - G$_{RP}$)$_0$, spreading over 4 magnitudes compared to a spread of 0.7 magnitudes in (G$_{BP}$ - G$_{RP}$)$_0$, as shown in Figure 2. The spread could be due to the age, metallicity, and extinction of the RC stars. As discussed further below, our optical CMD is similar to that of \citealt{gaiahrd} that shows the appearance of a RC for low extinction sources (E(B - V) $<$ 0.015) in DR2.

In Figures 1 and 2 we overplot our RC sample in the bottom panels. The RC stars show a clear NUV-optical color-dependent trend with [Fe/H] (higher [Fe/H] at redder (NUV - G)$_0$ and vice-versa, shown in Figure 1). This spread is unique to (NUV - G)$_0$ color, especially when the dust correction is applied. In Figure 3 we show the distributions for the RC stars of Galactic longitude, distance, NUV magnitude and E(B - V). The E(B - V) for most of this sample is less than 0.1, suggesting relatively small extinction corrections. Most of the RC sample is between 18 $<$ NUV $<$ 20. The GAIS survey limit is 21 (5$\sigma$) indicating this sample is reasonably complete out to our distance limit. 

Using the derived parameters from APOGEE, we show in Figure 4 that the RC UV-optical color also correlates with effective temperature. The trend for (G$_{BP}$ - G$_{RP}$)$_0$ is much weaker and more highly scattered. From a line fit, we measure $\sigma_{{BP-RP}_0}$ = 0.057 and $\sigma_{{NUV-G}_0}$ = 0.23. Compared to the slope of the RC in each color, the relative scatter is 2.6 times smaller in (NUV - G)$_0$ than in (G$_{BP}$ - G$_{RP}$)$_0$.

\section{Discussion}
In this section we define an ultraviolet-optical color-metallicity relation. First we separate our RC sample into two subsamples of thin and thick disk stars. Figure 5 shows [$\alpha$/Fe] vs [Fe/H] and distinguishes between thin and thick disk stars using a simple cut in [$\alpha$/Fe] (\citealt{hawkins15}, \citealt{nidever14}). The thick disk stars overall have a much higher [$\alpha$/Fe], especially at lower [Fe/H]. The [$\alpha$/Fe] - [Fe/H] relation is a unique way to separate different Galactic components to understand the star formation history of different parts of the Milky Way. The thin disk stars are thought to be considerably younger than the thick disk stars because of the smaller amount of $\alpha$ elements (O, Mg, Si, Ca, and Ti) for a given [Fe/H]. We see the same separation as in \citealt{nidever14} between the low and high $\alpha$ sequence around [Fe/H] = 0.2. 

The majority of thick disk stars are bluer and very likely hotter than the thin disk subsample. Alternatively, at a given (NUV - G)$_0$, the thick disk stars have a lower [Fe/H] than their thin disk counterparts. While none of the thin disk stars are metal poor enough to be considered halo stars (typically metallicities of [Fe/H] $<$ -0.5), they are not significantly different from the entire sample in other quantities except [$\alpha$/Fe]. The trends for the two different populations present in Figure 5 suggest different color-metallicity relations between them.

\begin{figure*}[]
\centering
\includegraphics[width=1\textwidth]{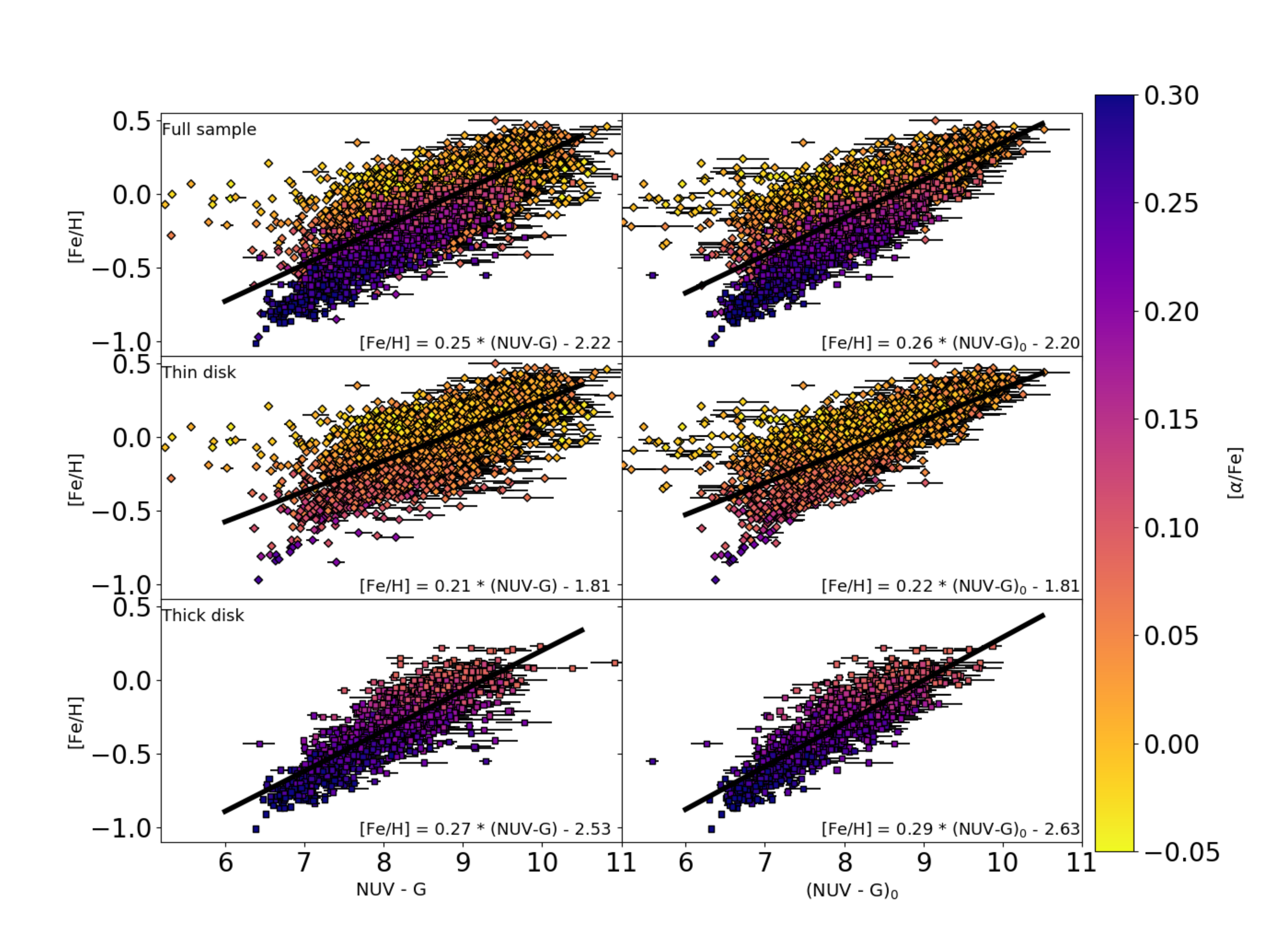}
\caption{[Fe/H] vs (NUV - G)$_0$ colored by [$\alpha$/Fe] without (left) and with (right) a Galactic extinction correction. We find several outliers at (NUV - G)$_0$ $<$ 6 that are possibly binaries. The middle panels show thin disk stars and the bottom panels show thick disk stars. Each fit is done only on the data in that panel. We include all six fits in their corresponding panels. The (NUV - G)$_0$ spread over 4 magnitudes is much greater than in optical colors. The color-metallicity relation has a weak dependence on [$\alpha$/Fe]. The trend tightens when corrected for Galactic extinction.}
\end{figure*}

In Figure 6 we show a clear trend between (NUV - G)$_0$ color and metallicity for the full sample that spans a wider range of color than in optical wavelengths as shown in Figure 1. This relation becomes much tighter when an extinction correction is added. We also separate the two thin and thick disk populations in the middle and bottom panels, respectively. We obtain the following relationship for the full extinction-corrected sample:

\begin{equation}
    [Fe/H] = 0.256 (NUV - G)_0 - 2.204.
\end{equation}

The standard deviation from the linear fit is $\sigma_{\rm{[Fe/H]}}$ $\sim$ 0.28 dex. Equation 1 provides a new means to determine metallicity from photometry with a precision similar to low-resolution spectroscopy (e.g. SDSS SEGUE, \citealt{lee2011}, where they measure [Fe/H] to a precision of 0.23 dex) but at a much cheaper cost and can be obtained for many more stars. 

The two different thin and thick disk populations from Figure 5 appear to have different color-metallicity relations. Thick disk stars have much less scatter from the relation than thin disk stars and the slope is higher in the thick disk color-metallicity relation. At the very metal poor end ([Fe/H] $<$ -0.6) RC stars appear to be a part of the galaxy's thick disk (\citealt{brook12}, \citealt{hawkins15}) and from Figure 5 they are bluer objects in general. The thin disk population also shows objects that are bluer than expected, including several outliers bluer than (NUV - G)$_0$ $<$ 6, some of which could be binaries. We will leave a more detailed discussion of these outliers to future papers. Even with the presence of outliers, the overall relation is still about as precise as spectroscopy. 

\begin{figure}[]
\centering
\includegraphics[width=0.5\textwidth]{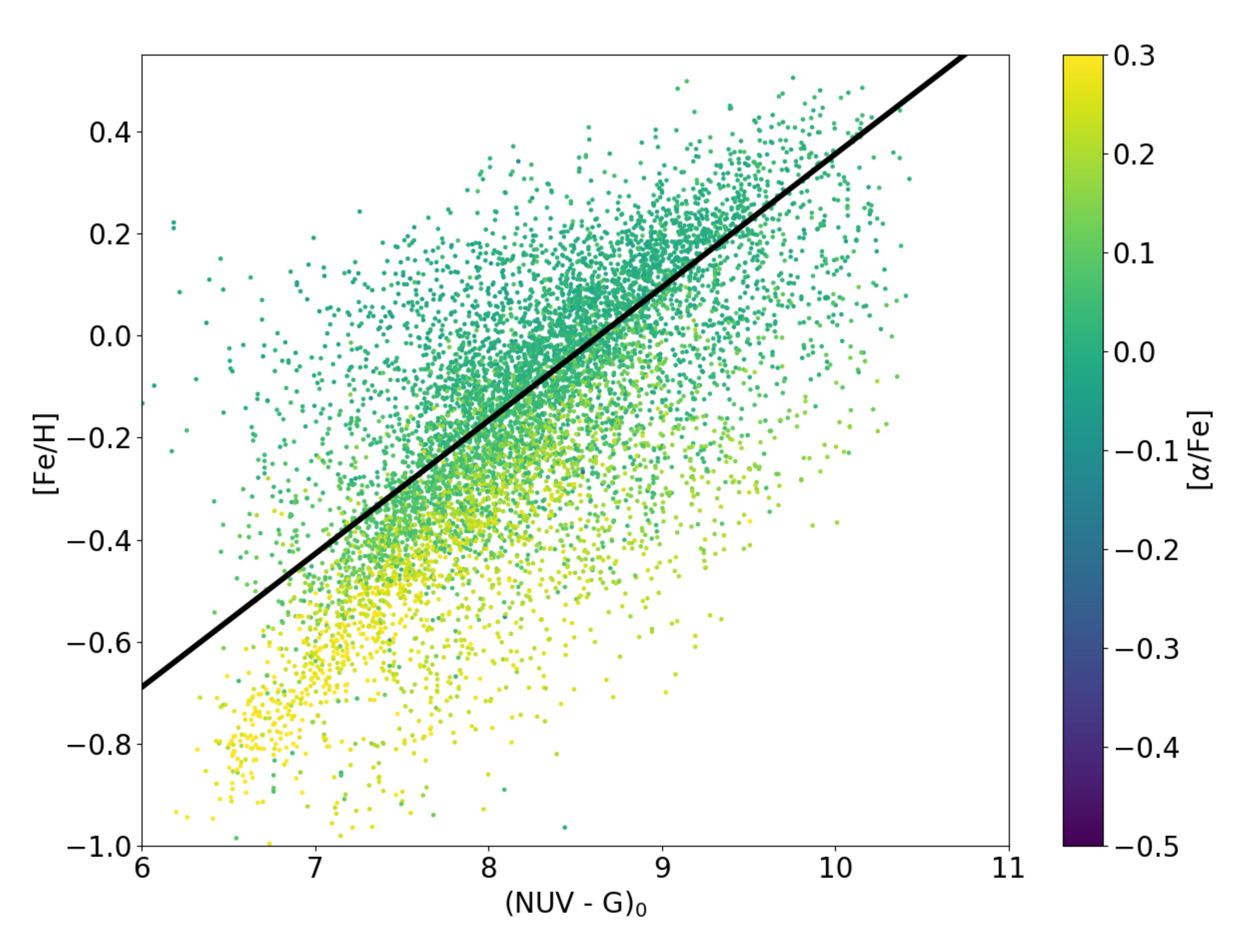}
\caption{[Fe/H] vs (NUV - G)$_0$ for stars within the RC box described in the text. The black line is the dust-corrected fit for the entire sample from figure 6, top right panel. The data match the trend we see at (NUV - G)$_0$ $>$ 7.5.}
\end{figure}

To demonstrate the effectiveness of this color-metallicty relation, we select a subset of photometrically-defined RC stars from the CMD in Figure 1. We define our RC box as bound at 6 $>$ (NUV - G)$_0$ $>$ 10.5 and 0.9 $>$ M$_G$ $>$ -0.1. This region contains many possible RC candidates from which one can also derive asteroseismic parameters with minimal contamination from RGB stars using the methods in \citealt{hawkins18}. The main sources of bias comes from dust extinction (and uncertainties) in NUV which would restrict the range of objects detectable by GALEX.

We apply the same cuts as described in section 2.2 to a matched catalog between GAIS, Gaia DR2, and APOGEE DR14 and replot the [Fe/H] - (NUV - G)$_0$ relation with these data. The overall trend between ultraviolet-optical color and metallicity closely matches that of the spectroscopically-obtained RC sample. The scatter in the relation is larger but this is likely due to contamination from other giant stars. There is an offset at (NUV - G)$_0$ $<$ 8 for the highest [$\alpha$/Fe] stars which likely is due to the steeper relation for the thick disk stars in this sample. 

Photometric-metallicity relations have been calculated or observed in the past. For example, \citealt{ivezic2008} use F and G main-sequence stars to derive a relation between [Fe/H], u - g and g - r. The u - g color, or the UV excess, depends on metallicity because of the high absorption of metals at bluer colors, affecting the star's flux. This UV excess depends on the g - r color which is related to the star's effective temperature. RC stars are known to have a flux-temperature relation that varies greatly depending on the metallicity. Metal line blanketing may also play a role due to the high [Fe/H] values in this sample (\citealt{girardi16}, \citealt{choi16}). Metals in stellar atmospheres absorb blue light due to metal line blanketing and should show significant absorption in bluer wavelengths like NUV. In our sample these metal-rich stars are the reddest, with (NUV - G)$_0$ values reaching up to 10 magnitudes versus the blue end at (NUV - G)$_0$ = 6 containing the most metal poor stars (Figure 6). Due to their similarity, these trends may also hold for stars along the giant branch albeit with increased scatter.

Finally we explore how our relation compares to predictions from a recently developed stellar evolutionary code. The MESA Isochrones and Stellar Tracks (MIST) model provides tracks and photometric outputs for a full range of stellar masses and metallicities with sufficient resolution to follow short-lived evolutionary stages (\citealt{choi16}). We used evolutionary tracks for the range of masses and metallicities likely to appear in the Milky Way field RC population, with  steps of [Fe/H] spaced by 0.25 dex and a uniform distribution of stellar masses. Luminosities were calculated in G, G$_{BP}$, G$_{RP}$, and NUV bands, using the most up to date Gaia bandpasses. We applied a photometric selection to the final outputs, selecting stars in a similar region for the RC in the CMD, as described above, while also restricting the model to the core helium-burning phase (MESA EEP 631-707).   In Figures 8 and 9 we plot NUV vs [Fe/H] colored by T$_{\rm{eff}}$ and log g, respectively. The extinction-corrected fits for the full sample and thick-disk-only subsample are shown on both plots.

The linear fits in Figure 8 provide a good match to the models over most of the metallicity range of our RC sample. Bluer stars tend to be hotter and have a lower metallicity and vice versa. The fit lines overlap models with the T$_{\rm{eff}}$ range in Figure 4.  Figure 9 shows that our mean fits are consistent with models with log g $\sim$ 2.4, with the thick disk stars having a slightly lower log g (and T$_{eff}$) at fixed metallicity. The larger scatter with increasing metallicity in the models is also seen in Figure 6, most notably in the thin disk subsample. This scatter can have many explanations (dust, binaries, star formation history). This topic will be the target of future papers.

RC stars are used as standard candles in infrared due to their constant absolute magnitude and color. Metallicity, mass, age, and extinction make their use as standard candles difficult in bluer wavelengths. Using the color-metallicity relation we can create a metallicity map of the Galaxy (e.g. \citealt{onal16}) and increase the accuracy of RC stars as standard candles. These results also have implications on the use of RC stars as extinction probes (see e.g. \citealt{girardi16}). \citealt{yanchulova17} use HST observations that extend to the NUV and explain the spread in color as due to extinction. They conclude the RC is confined to a small region in the CMD with similar metallicities. We would expect to see a large spread in (NUV - G)$_0$ over a comparable metallicity range, indicating that metallicity may play a non-trivial role in understanding the RC in CMD space.

\begin{figure}[]
\centering
\includegraphics[width=0.4\textwidth]{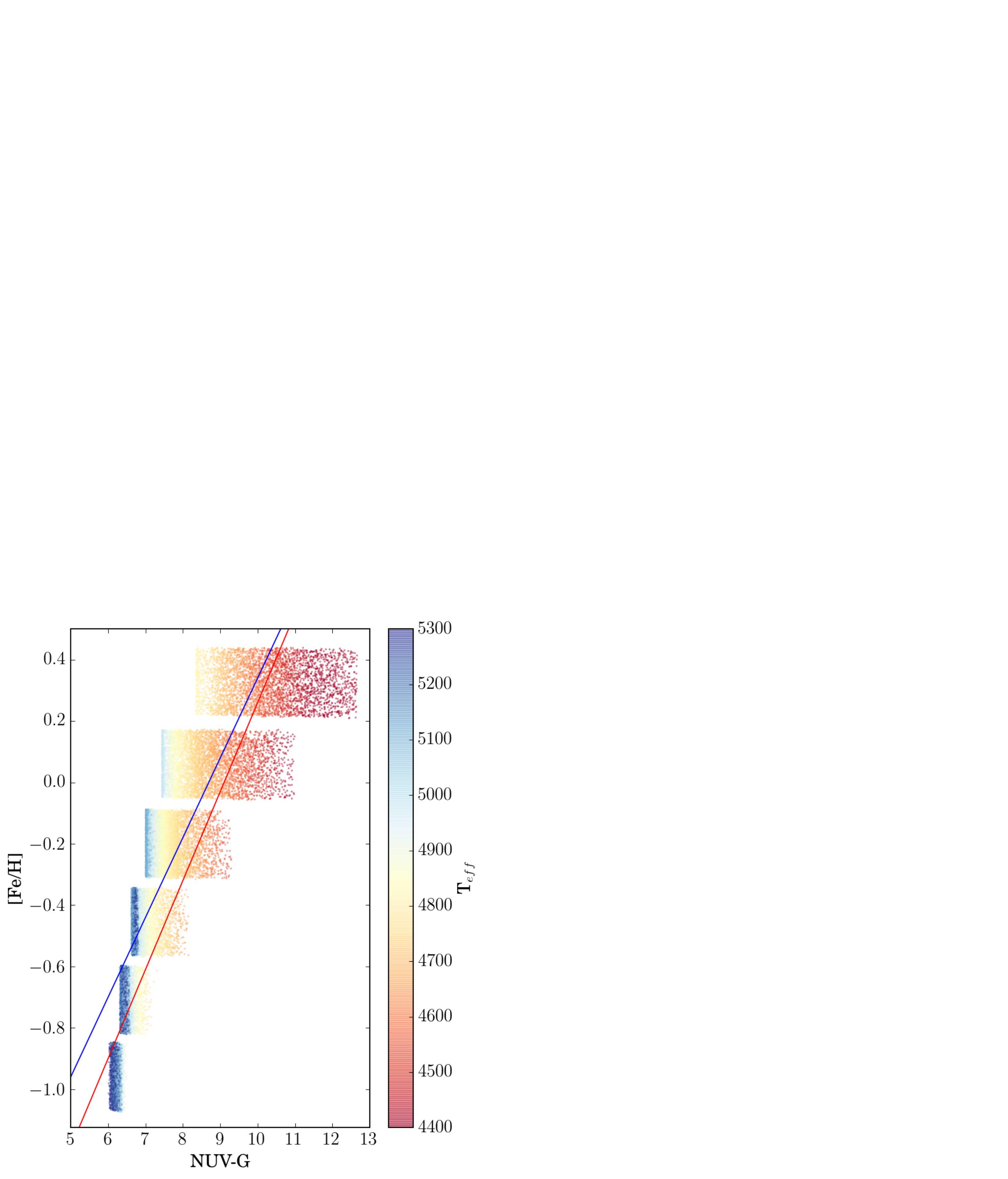}
\caption{Figure 6 plotted with MIST stellar evolution models colored by T$_{\rm{eff}}$. The blue line is the fit to the entire sample and the red line is the fit to only the thick disk.}
\end{figure}

\begin{figure}[]
\centering
\includegraphics[width=0.4\textwidth]{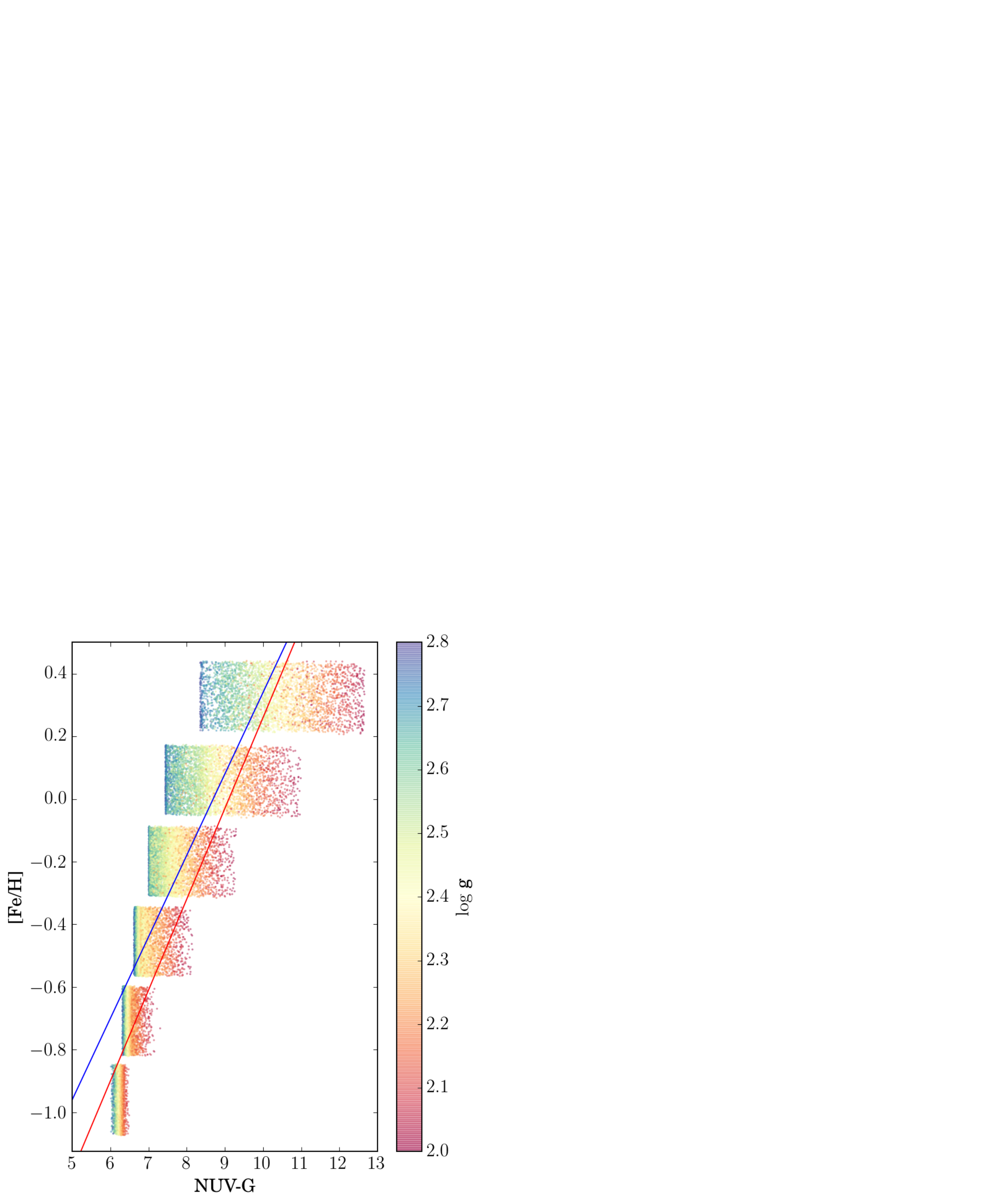}
\caption{Figure 6 plotted with MIST stellar evolution models colored by log g. The blue and red lines are the same as in Figure 8.}
\end{figure}

\begin{deluxetable*}{ccccccccccc}
\tablehead{\colhead{RA} & \colhead{Dec} & \colhead{NUV} & \colhead{(NUV - G)$_0$} & \colhead{G$_{BP}$} & \colhead{(G$_{BP}$ - G$_{RP}$)$_0$} & \colhead{E(B - V)} & \colhead{DM} & \colhead{[Fe/H]} & \colhead{T$_{\rm{eff}}$} & \colhead{[$\alpha$/Fe]} \\ 
\colhead{(\deg)} & \colhead{(\deg)} & \colhead{(mag)} & \colhead{(mag)} & \colhead{(mag)} & \colhead{(mag)} & \colhead{(mag)} & \colhead{(mag)} & \colhead{} & \colhead{(K)} & \colhead{} } 
\startdata
0.1300 & 15.2717 & 18.11 $\pm$ 0.03 & 7.46 $\pm$ 0.03 & 11.17 $\pm$ 0.00 & 1.16 $\pm$ 0.00 & 0.04 & 10.41 & -0.41 $\pm$ 0.01 & 4931.00 & 0.10 $\pm$ 0.02 \\
0.2081 & 16.3654 & 19.40 $\pm$ 0.07 & 8.82 $\pm$ 0.07 & 11.18 $\pm$ 0.00 & 1.31 $\pm$ 0.00 & 0.04 & 10.13 & 0.06 $\pm$ 0.01 & 4699.00 & 0.01 $\pm$ 0.01 \\
0.3194 & 15.3927 & 19.49 $\pm$ 0.06 & 9.78 $\pm$ 0.06 & 10.36 $\pm$ 0.00 & 1.39 $\pm$ 0.00 & 0.04 & 9.10 & 0.34 $\pm$ 0.01 & 4553.00 & 0.02 $\pm$ 0.01 \\
0.4016 & 0.2359 & 18.97 $\pm$ 0.04 & 8.80 $\pm$ 0.04 & 10.74 $\pm$ 0.00 & 1.26 $\pm$ 0.00 & 0.02 & 9.61 & 0.04 $\pm$ 0.01 & 4723.00 & -0.01 $\pm$ -0.42 \\
0.5204 & 15.0377 & 19.08 $\pm$ 0.05 & 7.55 $\pm$ 0.05 & 12.04 $\pm$ 0.00 & 1.17 $\pm$ 0.00 & 0.05 & 11.27 & -0.36 $\pm$ 0.01 & 4915.00 & 0.13 $\pm$ 0.02 \\
0.6003 & 16.4273 & 17.30 $\pm$ 0.02 & 7.94 $\pm$ 0.02 & 9.90 $\pm$ 0.00 & 1.20 $\pm$ 0.00 & 0.03 & 9.03 & -0.24 $\pm$ 0.01 & 4826.00 & 0.03 $\pm$ 0.02 \\
0.6012 & 16.9140 & 19.10 $\pm$ 0.06 & 8.48 $\pm$ 0.06 & 11.18 $\pm$ 0.00 & 1.24 $\pm$ 0.00 & 0.03 & 10.17 & -0.01 $\pm$ 0.01 & 4782.00 & 0.03 $\pm$ 0.06 \\
0.6975 & 17.1347 & 17.58 $\pm$ 0.03 & 8.38 $\pm$ 0.03 & 9.77 $\pm$ 0.00 & 1.24 $\pm$ 0.00 & 0.03 & 9.25 & -0.23 $\pm$ 0.01 & 4734.00 & 0.02 $\pm$ 0.01 \\
0.9191 & 16.9788 & 18.68 $\pm$ 0.05 & 8.95 $\pm$ 0.05 & 10.37 $\pm$ 0.00 & 1.37 $\pm$ 0.00 & 0.03 & 9.18 & 0.17 $\pm$ 0.01 & 4548.00 & 0.05 $\pm$ 0.04 \\
0.9634 & 75.8578 & 19.44 $\pm$ 0.34 & 8.37 $\pm$ 0.34 & 11.79 $\pm$ 0.00 & 1.53 $\pm$ 0.00 & 0.30 & 10.67 & 0.15 $\pm$ 0.01 & 4929.00 & -0.01 $\pm$ -0.01 \\
\enddata
\tablecomments{(1) Gaia RA, (2) Gaia Dec, (3) GALEX NUV, (4) dust corrected NUV - G, (5) Gaia G$_{BP}$, (6) dust corrected G$_{BP}$ - G$_{RP}$, (7) E(B - V) from \citealt{GSF15}, (8) Distance Modulus, (9) stellar metallicity, (10) the effective temperature from APOGEE $\pm$ 91.47 K for all values, (11) alpha abundance. Table 1 is published in its entirety in the machine-readable format. A portion is shown here for guidance regarding its form and content.}
\end{deluxetable*}

\section{Conclusion}
Using a sample of 5,175 RC stars from APOGEE with data from GALEX and Gaia, we identify the RC in UV-optical CMD space as well as the existence of a color-metallicity relation that is tighter in (NUV - G)$_0$ than (G$_{BP}$ - G$_{RP}$)$_0$.  We see a strong dependence of color on T$_{\rm{eff}}$ and metallicity. As part of this analysis, we apply a Galactic extinction correction using a 3-D dust map from \citealt{GSF15} and Gaia distances, which further tightens the relation. If we separate the sample into thin and thick disk stars, thick disk stars appear bluer than their thin disk counterparts for a given temperature and redder at a fixed metallicity. Finally, we find a tight relation between (NUV - G)$_0$ and [Fe/H] with a standard deviation of about $\sigma$ = 0.28 that can be used to estimate stellar metallicities of RC stars when a spectroscopic metallicity measurement is missing. This relation will be used to obtain photometric metallicities from other stars in the same CMD space as RC candidates using only their UV-optical color. 
An NUV GALEX Plane Survey (Mohammed et al., in prep) will provide NUV measurements for the Galactic Plane for the first time using GALEX. This survey will provide millions of new objects brighter than NUV = 20 magnitude that will aid in RC investigations as well as many other fields in Galactic astronomy. Spectroscopic followup of these candidates could confirm their RC status using the method of \citealt{hawkins18} and \citealt{ting18} and allow us to further understand the UV-optical color-metallicity-age relation. RC stars are excellent extinction probes and if their metallicity is known it is enough to use a CMD to fit for its extinction values. Using our color-metallicity relation in conjunction with extinction measurements from \citealt{GSF15} we can narrow the variables to mass and age. 

\acknowledgments
We acknowledge support from NASA-ADP Grant NNX12AI50G. GALEX ($Galaxy$ $Evolution$ $Explorer$) is a NASA Small Explorer, launched in 2003 April. This work is based on data from the European Space Agency (ESA) mission Gaia (https://www.cosmos.esa.int/gaia), processed by the Gaia Data Processing and Analysis Consortium (DPAC, https://www.cosmos.esa.int/web/gaia/dpac/consortium). This study makes use of the publicly released data from APOGEE DR14. This research made use of Astropy, a community-developed core Python package for Astronomy (\citealt{astropy}). We acknowledge Dustin Lang for timely production of the Gaia DR2-GALEX catalog.

\bibliographystyle{aasjournal}
\bibliography{ms.bbl}

\end{document}